# User Data Sharing Frameworks: A Blockchain-Based Incentive Solution

Ajay Kumar Shrestha
*Department of Computer Science*
*University of Saskatchewan*
Saskatoon, Saskatchewan, Canada
ajay.shrestha@usask.ca

Julita Vassileva
*Department of Computer Science*
*University of Saskatchewan*
Saskatoon, Saskatchewan, Canada
julita.vassileva @usask.ca

*Abstract—* **Currently, there is no universal method to track who shared what, with whom, when and for what purposes in a verifiable way to create an individual incentive for data owners. A platform that allows data owners to control, delete, and get rewards from sharing their data would be an important enabler of user data-sharing. We propose a usable blockchain- and smart contracts-based framework that allows users to store research data locally and share without losing control and ownership of it. We have created smart contracts for building automatic verification of the conditions for data access that also naturally supports building up a verifiable record of the provenance, incentives for users to share their data and accountability of access. The paper presents a review of the existing work of research data sharing, the proposed blockchain-based framework and an evaluation of the framework by measuring the transaction cost for smart contracts deployment. The results show that nodes responded quickly in all tested cases with a befitting transaction cost.**

I. INTRODUCTION

The internet from its inception was aimed to facilitate users in sharing data, and it enabled it through centralized (e.g. FTP) or decentralized (e.g. email) services. With Web 2.0 [1] or the social web, it became very easy to share creative products on social sites (user-generated content on YouTube, Wikipedia, blogs, as well as microblogging tools like Twitter and Facebook). To achieve profit in the business model, secondary data associated with the users' profile and behaviour are being collected and shared among the enterprises, for a personalized advertisement that targets the users based on that information. Much of the data are contributed voluntarily by the user; others are obtained by the system from the observation of user activities or inferred through advanced analysis of volunteered or observed data [2].

In different domains such as tourism, e-commerce, news aggregators, dating services etc., the data analytics and the personalization enhances the users' interaction with the system, and the overall quality of services being offered to the users. The applications that aim for personalization need to gather information about their users and create predictive user models, used to adapt their functionality, presentation, or offers to the specific users' requirements [3]. The process of user modeling requires collecting user data and making inferences from this data by both finding patterns and similarities across the many users of a service, or by abstracting user features and building user profiles from the history of the interaction of a user [4]. This is a slow process prone to the "cold start" problem, and its variants (new user, or new product/feature) [5]. To speed up learning about their users, applications can share relevant data about the same user with other applications, leading to the need for sharing user interaction data and user profiles [6]. This sharing is very problematic ethically, and the recent EU's General Data Protection Regulation (GDPR) makes explicit the problems and trade-offs related to user privacy and control over data, as well as fairness while preserving the richness of data. Our research shows that these problems have not been addressed by different proposed architectures and methods for user profile data sharing.

Most importantly, in the scientific research domain, research data sharing practices are much needed to maximize the knowledge gains from the research efforts of millions of researchers. Sharing research data can reduce duplicative trials and accelerate discovery. In medicine and healthcare, both personalized patient care and medical research can benefit from ethical and privacy-preserving sharing of patient data and data from clinical trials [7]. A flexible mechanism for obtaining and renewing consent for data use and sharing is required that provides appropriate and meaningful incentives to capitalize from data sharing and ensures transparency for users to be aware of which of their data has been accessed, by whom, for what purpose and under what conditions.

Currently, there is no single trusted authority to ensure ethical user data sharing. It has been demonstrated that the creativity and the advancement of the technologies have given birth to many computational backbones to ensure privacy and data sharing model that include cloud computing services, intelligent computing etc. However, these services are often criticized due to security and centrality issues, and for the credibility of the services being offered. When the trust resides within a centralized service provider for all the storage and management of data, it presents a single point of failure, and attractive target for hackers [8], creating risks of data being stolen and misused, or sold to unethical third parties, or even destroyed, if the centralized service goes bankrupt.

In the past few years, distributed ledgers and blockchain technology have evolved as a promising means to support immutable and trusted records in various use cases including healthcare, agricultural research, tourism domains etc. In addition, many Blockchain systems provide a technology called "smart contract" that allows building automatic verification of the conditions for access or modification of each data entity. Smart contracts can be deployed to encode allowed purposes of data use, allowed software apps, people or businesses who can access the data, time limitations, price for access, etc. Therefore, the blockchain provides a new type of platform that is useful for sharing user data - both user models and user-contributed data or research data, by providing solutions to the privacy, user control and incentives problems. The most important criticisms to blockchain-based approaches relate to their performance. However, the current state of development of the technology allows, through thoughtful combinations of blockchains to achieve acceptable performance, as we show in this paper.

In the next section, we present some background on blockchain and smart contracts. Sections III presents our definition of user data and its three distinct categories. A review of recent developments in user-controlled access and sharing of research data is presented in section IV. Section V details our blockchain-based solution and section VI presents the testing and evaluation of our approach. Section VII concludes the paper with future research directions.

## II. BACKGROUND

### A. Distributed Ledger Technologies

Distributed ledger technology (DLT) is a data structure used to create a public or private distributed digital transaction ledger, which, instead of resting with a single provider, is shared among a distributed network of computers. Several variations of DLTs such as Hashgraph [9], NANO (formerly known as RaiBlocks) [10], Peaq [11] etc. have been created to support the continual growth and change of the distributed crypto-world. However, blockchain technology holds promise to transform data management and business models in many domains. Initially, blockchain was also considered for only powering virtual currency [12], but the applications of the blockchain technology have since quickly evolved to numerous use cases [13]. The basic software pattern of blockchain was introduced in the original source code for the digital cash system, Bitcoin [14] and implemented in 2009 by mining it for the very first time.

Fig. 1 represents a blockchain consisting of a sequence of blocks. The block in the blockchain is composed of several transactions, which depends on the size of the block [15]. The block contains vital information in its block headers such as previous block's hash, which points to the previous block, Merkle tree root hash, which is the aggregate hash value of all the transactions' hashes in the block, and timestamp, which records the time in the UNIX Epoch time.

The key idea is that, computationally, it is impracticable for any corrupted node (unless the number of such nodes is higher majority consensus) to go back and alter the history. There is no single point of failure in blockchain because the redundancy of the system ensures many backups, and the lack of a central storage place ensures there is no one target for hackers.

To ensure the success of the blockchain platform, incentives play a very important role in encouraging participation. Usually, the nodes are incentivized with mining rewards for taking part in the transaction validation activities, such as bitcoin/ Ethereum miners. Additionally, there are some proposals for using blockchain incorporating access control measures to ensure the privacy of data while sharing off-chain files [16], [17]. User modeling can benefit from a platform using distributed ledgers and smart contracts to ensure user-controlled privacy and data-sharing policies encoded in smart contracts. In contrast to the centralized system, blockchain technology can be transparent to the users and very promising to incentivize users for data sharing. It also naturally supports building up incentives for users to share their data, in terms of rewards (micro-payments or credits) encoded in the smart contracts. In this way, users become owners of their data and can decide how their data is collected and used, as well as shared, and benefit not only in terms of improved personalized experience with the service but also directly, for example, by participating in the share of the advertising revenue generated by the service provider.

### B. Smart Contract

Research in the smart contract technology has evolved from the conceptual-based architecture to the application-oriented scenarios. According to the systematic mapping study conducted by Alharby et al. [18] in 2018, 64% of total 188 relevant papers on smart contracts were from the applications category compared to a very few application-oriented academic papers in 2017, when they extracted only 24 papers in total. Their 2017 study shows that about 66% of the papers focused on the conceptual level of finding and tackling smart contract issues. We have now seen many academic researchers taking up smart contract technologies in actual building applications on the top of the blockchain.

The smart contracts are now recognized as the instances of contracts deployed on the Ethereum blockchain [19] although it was originally coined in the paper [20] to design the electronic commerce protocols between strangers on the Internet. A Smart contract stores the rules which (a) negotiate the terms of the contract, (b) automatically verifies the contract and (c) executes the agreed terms. A smart contract consists of different functions that might be called from outside of a blockchain or by other smart contracts. Blockchain coupled with smart contract technology removes the reliance on the central system between the transaction parties. Since the smart contracts are stored on the Blockchain, all the connected parties in the network will have a copy of them.

The Solidity programming language [21] is used to write smart contracts because it only allows performing basic operations on its basic data types resulting in lightweight code. The Ethereum Virtual Machine (EVM) code [19] is used in the contracts and that consists of bytes, each representing an operation. The code can access the amount of Wei sent in the transaction and data of the incoming message, block header data, and return a byte array of data as an output. Wei is the smallest denomination of ether in the Ethereum network (1 Wei = ether/1e18). With the implementation of Ethereum Web Assembly (EWASM) [22] in the near future, smart contract development can be done in any other programming languages besides Solidity and that will speed up the function call between Web Assembly and JavaScript.

## III. USER DATA

In our research work, we have categorized user data into three types: (1) User profile data, (2) User documents, and (3) Research data.

The actual characteristics of the user are defined as the user profile data, which shows who the users really are. According to [23], the user profile information can be collected in real-time through the app, or outside the app and imported to the analytics platform so that the applications can offer personalized services to the users.

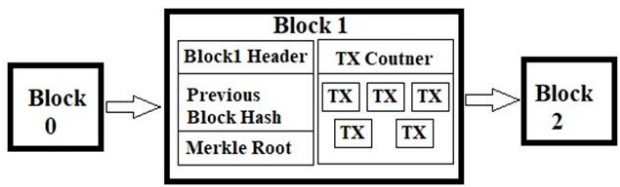

Fig. 1. An example of blockchain

User documents are the user-generated content (UGC) or user-created content (UCC), which can be represented in different media, such as text, videos, audio, photos. They evolved primarily in the business world with many discussions around free content for organizations and got recognition with the emergence of participative web or "Web 2.0". UGC and UCC both refer to the same content, as stated in [24] which is "a generic term comprising a wide range of media and creative content types that were created or at least substantially co-created by contributors (users) working outside of conventional professional environments". Therefore, we have presented a common new term "user documents" to indicate UGC or UCC.

The last category of user data is research data, which can be user profile data or user documents, or any factual records born during the course of conducting a research activity and used as primary sources for analysis to produce original research results. The research data cannot be private, rather considered as "owned" by the researcher or "public" based upon the rules within the data governance policy under which the data was collected. These rules are often deliberately vague about the outcomes of the research. The research data has always been a way to validate research outcomes and can be combined with other information to accelerate new research findings. Most of the time, the research data are collected, observed and recorded, or created in digital format such as spreadsheets, images, videos, survey data, experimental data, sensor data, artifacts, scripts etc. which make them easier to share. In this paper, we propose a blockchain-based solution for sharing research data.

## IV. SHARING RESEARCH DATA

Sharing research data enable researchers to start a new collaboration. However, the ownership of the data as an important asset to the researcher, in a competitive research environment creates negative incentives for sharing. Most researchers, on an individual level, may feel reluctant to share their research data, however, they appreciate the overall benefits of data sharing, which was also concluded from the qualitative interviews-based study conducted in [25], [26]. Those studies also recognized six different ways of data sharing such as private management sharing, peer exchange, community sharing, collaborative sharing, sharing for transparent government and public sharing. The original researchers must feel motivated towards sharing their research data since many data are a significantly important and valuable resource beyond usage for them. We have compiled some of the important motivations or influential factors for the original researcher to share research data. They are as follows:

- Sharing research data is an integral part of the research activity.
- The standard guidelines, policies, data services and research disciplines that they follow.
- Transparency, accountability, scrutiny of research outcomes.
- Expectations of funder and publisher.
- Increase in the research's impact and its visibility.
- New collaboration can lead to their career benefits.
- Direct credits and attributions are received for their efforts in collecting the data.
- The duplicate research trials are reduced.
- Validation of research methods is encouraged.
- Educational and training materials can receive resources.

Proper data organization is equally important not just for original researchers to validate their research results but also to facilitate other researchers for accessing the research data and working on them. Although research data at different stages have a different structure, researchers must maintain the minimum documentation such as the organized version control commits in order to keep track of the collected data and ascertain that they are usable. Besides, the documentation must also include the overall structure of the research dataset, relationships between various data files and information on data confidentiality. Moreover, the mechanism for querying data and providing answers rather than full sharing of data is very much needed. However, current services are missing these characteristics since the data is not uniform and it would generally require complex and difficult processes [27] to provide such services.

We have categorized different mechanisms for sharing research data into three distinct groups, in terms of the level of control, degree of openness and availability of research data, which are:

- Centralized,
- Decentralized,
- Blockchain-based (decentralized + incentives).

The centralized notion indicates that the research data should be kept in only one database file, located at a single point at a given time period on a given network and the control of data is exerted by just one entity. The single authorized entity manages the central filesystem that must be capable of maintaining all the received data from the researchers and responding to every single query coming from different agents by themselves. Mostly the centralization in research data sharing is non-distributed with the same proprietary physical storage for the data files. Examples include universities, funding agencies depositing research data into their institutional repository or submitting research data to a journal to support a publication.

On the other hand, the underlying idea of the decentralized dimension is that the research data can be stored in multiple physical locations at a given time on any network, and the control is shared among various entities. The distributed storage strategy gives resiliency and possesses a high degree of availability. The researcher can use cloud services, web/email services or off-line USB stick to share research data files to other interested agents, who contribute to the replication of the data in a different physical location.

In the context of a blockchain-based data-sharing solution, MedRec [28] uses blockchain technology for the first time to preserve the privacy of user data while handling electronic medical records (EMRs) in a completely distributed P2P network. The participating medical stakeholders (researchers, public health authorities) in the network act as blockchain "miners". The system incentivizes the participants by giving them access to aggregate, anonymized data as mining rewards, in return for sustaining and securing the network as miners. However, it only collects static data from a medical examination record and will be inefficient to support metadata change while sharing data streams generated continuously from sensors and other monitoring devices.

For data sharing in the scientific research domain, Shrestha & Vassileva [17] provides a usable blockchain-based model for a collection of researchers' data, providing accountability of access and incentives for sharing, maintaining the complete and updated information, and a verifiable record of the provenance, including all accesses/sharing/usages of the data. Besides, the work in [16] provides an off-chain storage solution along with a decentralized personal data management model using blockchain that allows the data owner to share secret keys to the data requester via some secure channel such that only the users holding secret keys can access the data. However, it has high communication overhead leading it to be practically infeasible.

In [29], the authors mentioned some of the challenges present in the proposals adopting Blockchain as part of their solutions. The authors discussed that the high cost of having all the operations performed over the Blockchain makes them impractical. So, all the operations that aren't crucial to be in the Blockchain should be moved away from it and directed to the Trusted Third Party (TTP), which is a gateway assistant giving insights about allowing or denying attempts of data access. We have adopted this suggestion in our solution.

Many technology experts have suggested that the blockchain will make internet decentralized and become a crucial component of the next generation of the internet [30]. In the context of decentralizing the internet, Ethereum blockchain provides Ethereum Virtual Machine (EVM) as a sandboxed virtual stack on each of the Ethereum full nodes that enable enterprise decentralized applications- dapps to run on the decentralized network. Ethereum is the world's second-biggest blockchain supporting smart contracts with which users can code, deploy and execute their contracts to deal with assets such as user data, commodities or goods in the supply chain marketplace. Ethereum has as its own virtual currency called '*ether*' (ETH), which is used to pay a transaction fee and provide a primary liquidity layer for exchanging digital assets.

## V. PROPOSED BLOCKCHAIN-BASED SOLUTION

Some of the desirable characteristics of an ideal system to enable users to store and share research data are:

- Rewards for honesty: An incentive mechanism that provides rewards to each of the participating entities (data owner; data producer, data provider and data consumer) for doing their part honestly or penalties for their malicious behaviors, in such a way that rational parties be persuaded to be truthful.

- Authorization: The system should define access policies, specify access rights and privileges to each party involved in the system.

- Data deletion: The system should comply with the EU's GDPR to identify and localize all personally identifiable information (PII), categorize data into PII and non-PII, apply data deletion rule for each category of data.

- Integrity: None of the participants is authorized to alter the audits and the agreed terms and conditions.

- Auditability: The system must provide the complete audit features to trace back data's state and route. It should be possible to track every action performed by the participants.

Therefore, we present the proposed solution based on blockchain from a performance engineering perspective, with an emphasis on access to the data stored off-chain. The blockchain solution is to manage the access control rather than the storage of the data. By design, our solution inherits key features of the blockchain such as immutability, decentralized trust, non-repudiation, and availability. To ensure privacy, our solution doesn't give direct access to data to any other party but provide query-based access, where the smart contracts regulate what kind of transformation is done to the data, to preserve privacy according to the wishes of the user. We present a general blockchain-based approach and identified the challenges across the different layers of blockchain systems, specifically on the off-chain data storage in this section. Then, we provide the implementation of our solution that utilizes the smart contracts to incentivize data owner for sharing their data and ensure honesty from all the participants in the system by imposing a collateral deposit.

### A. System Design

The privacy-related legislation such as GDPR regulates the processing of the personal data of the individuals and demands personally identifiable data erasure. However, data on the blockchain are always immutable. This could be a challenge while adopting blockchain as part of the solution. Therefore, we have a general architecture of the user data sharing system based on blockchain with off-chain data storage as shown in Fig. 2.

The general architecture ensures that the actual user data is never exposed to the blockchain. The user data is first hashed and encrypted before uploading into the off-chain storage. The data owners from their client applications can directly store them on the off-chain storage or alternatively use the trusted program as a gateway assistant to do so. The trusted program runs on a cloud in a trusted execution environment (TEE) [31] and mainly provides key management service for users and executes off-chain data operations that were negotiated between two parties- data owner and data consumer.

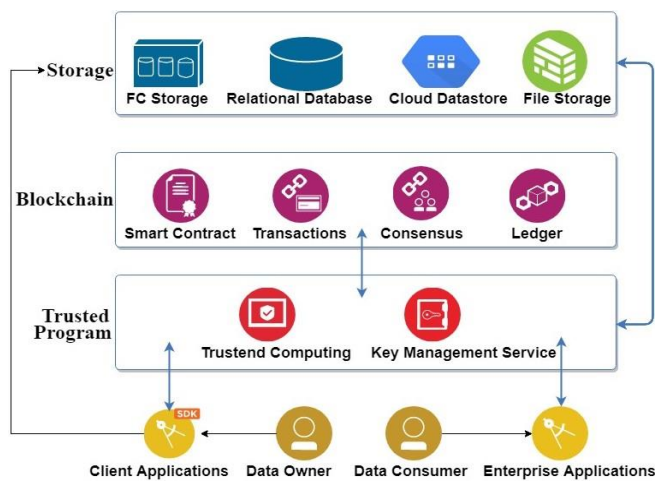

Fig. 2. General System Architecture

The terms and conditions regarding the access to user data are encoded in smart contracts along with the metadata and hash of the data and published on a blockchain platform (Ethereum). The hashes of the data on the blockchain prevent the middleware from tampering the data. The content-based addressing makes hashes of data serve as their identifier for retrieval. When the data consumer invokes the smart contracts for accessing the user data, only the successful invocation of the contracts results in the release of the key for decrypting the user data. The trusted program then extracts the hash from the blockchain, uses this hash to retrieve the data from the off-chain storage, decrypts it and releases the data to the data consumer while settling the incentives for the data owner. Blockchain and smart contracts support users by giving the users full transparency over who accesses their data, when and for what purpose, allowing the users to specify a range of purposes of data sharing, kinds of data that can be shared, and classes of applications/companies that can access the data, and providing an incentive to the users for sharing their data (in terms of payment for the use of the data by applications, as specified by the contracts).

The general architecture presents the underlying off-chain user data storage mechanism which could be a centralized data store hosted by a trusted party or decentralized platform such as InterPlanetary File System (IPFS) [32]. The centralized storage at worst case could be affected with centrality issues such as intentionally deleting the user data or not delivering the user data due to a technical failure. Therefore, the IPFS based storage with its decentralized delivery mechanism offers a better system where participating nodes don't have to rely on any individual system to obtain the data. However, the IPFS also doesn't offer a built-in data encryption scheme for the blockchain system.

### B. System Implementation

We present a framework with separate private permissioned MultiChain as a solution to both on-chain and off-chain data storage, encryption, hashing and tracking of data, together with Ethereum (for access control). Off-chain blockchain implementation with user data storage can be successfully achieved with the limited number of peers in the MultiChain [33]. It has the potential to replace the traditional centralized databases used to store user data into decentralized manner, offering more cryptographic auditing feature. Users can optionally store any published data in off-chain that saves storage place and bandwidth. It can hash off-chain data into the blockchain, with the data itself, delivered rapidly over the P2P network. The similar idea of storing off-chain data with MultiChain blockchain has also been proposed in research works by [34], [35]. MultiChain provides the privacy and control required in an easy to configure and deploy package [33]. It supports UNIX and Windows servers and comes up with a rich JSON-RPC API for easy integration with existing systems.

Fig. 3 presents the interaction among the data owner (provider) and data consumer of the proposed blockchain-based user-controlled data sharing scheme. This framework is based upon our general architecture with some modification, as the data owners, in this case, use the MultiChain for the on-chain and off-chain storage of their data. MultiChain nodes handle key operations such as hashing and encrypting the user data, storing the encrypted file locally (outside of blockchain), committing the hash of the file on the blockchain, searching the required data, verifying the data and delivering the data.

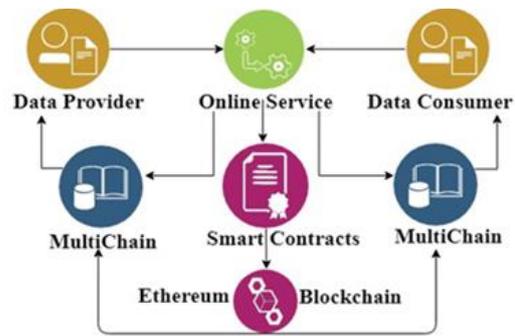

Fig. 3. User-Controlled Data-Sharing Framework

All the participants in the system with their Ethereum account addresses are in the MultiChain network. The data owner puts data in the local storage. The MultiChain enables to encrypt it and slice the large dataset into smaller segments. It then creates a transaction with their hashes and commits it into the blockchain. The data consumer node subscribes to the stream searching for the items (data) and finds the off-chain item with the help of their metadata and hashes. It then places the hash portion in its retrieval queue that queries the data in the P2P network. The node which possesses the data signature (data owner) responds to the query. At this point, the smart contracts get triggered and with their successful execution, the tokens are transferred from the data consumer's Ethereum address to the data owner's account while delivering the requested data (with verified hashes) to the local storage of the consumer node using the same path.

To support confidentiality, three blockchain streams with a combination of symmetric and asymmetric cryptography have been used. One is Pubkeys stream, which is used by participants to distribute their public keys under the RSA public-key cryptography. The second one is Items stream, used to publish large pieces of data, each of which is encrypted using a symmetric AES cryptography scheme. And the last one is Access stream which provides data access. For each participant who should see a piece of data, a stream-entry is created which contains that data's secret password, encrypted using that participant's public key.

We have created smart contracts for building automatic verification of the conditions for access of each data entity that also naturally supports building up incentives for users to share their data, in terms of rewards (micro-payments or credits). The smart contract imposed double deposit collateral to ensure that all participants act honestly. The steps to generate the codes are given in Algorithm 1.

| ALGORITHM 1: confirmIncentivesForSharingData |
|---|
| Input: deposit, dataPrice, contractState, $A_{cu}$, $A_{co}$, $A_{cc}$: Set of Ethereum addresses of data owners, data consumers and contracts respectively. |
| 1. Grant access to only $a_{cu} \in A_{cu}$, $a_{co} \in A_{co}$ who got registered into the system. |
| 2. Change the contract state to Created. |
| 3. $a_{cu}$ deposits $e_d$. |
| 4. Set data price to $e_p$ such that $e_p = \frac{1}{2} e_d$. |
| 5. Contract balance of $a_{cc}$ is $r_b = e_d$, where $a_{cc} \in A_{cc}$. |
| 6. Allow $a_{co}$ to choose the customer data of its interest. |
| 7. $a_{co}$ decides to consume the customer data $a_{cu}$, pays $2e_p$ such that consumer's deposit = $e_p$. |

8. Contract balance of $a_{cc}$ is $r_b = e_d + 2 * e_p$.
9. Change the contract state to Locked.
10. Grant $a_{co}$ to access the customer data.
11. $a_{co}$ confirms data availability
12. Change the contract state to Inactive.
13. Transfer deposit $e_p$ from $a_{cc}$ back to $a_{co}$.
14. Remaining Contract balance of $a_{cc}$ becomes $r_b = e_d + e_p$.
15. Transfer $r_b$ to $a_{cu}$.

Whenever a contract is published, it becomes a part of the transaction on the blockchain and possesses an address. Every transaction has a message which is a container object that can keep metadata. So, we have also created another simple contract to commit the ID and metadata related to the research into the blockchain. This resolves the issue of proof of the existence of the research activities for the researchers. We provide the user-transparent contracts with the help of state, functions, modifiers and events of the Ethereum smart contracts. We have setter functions to change the state of the contracts, whereas getter functions for all the public state variables are created implicitly. We have functions for depositing collateral amount, adding research provenance into the blockchain, settling payment for trading research data. The modifiers are implemented to restrict the access to calling only particular functions by confining the smart contracts' functions as per the roles of the participating entities to execute various operations between the data owner and consumer. There is also event associated and called in every function to write a log entry into the block as a part of the transaction so that the logs can be traced back later during disputes. The logs are also useful for decentralized applications by tying them into JavaScript callback functions because Ethereum JavaScript SDK can listen to every event.

In the case of the complex contract codes demanding a huge amount of data and computation, we can take out some computational parts out of the smart contracts and execute them in a trusted execution environment, which is also governed by the smart contracts on the chain. The attestation of the off-chain trusted execution environment (e.g. Intel SGX) could be done remotely to establish trust once the permission was granted by the on-chain smart contract. However, it is still a challenging part to ensure the credibility of the contract related off-chain executions in the trusted execution environments outside of the blockchain, as there have been some recent research findings to disclose the flaws of the SGX platform [36]. Therefore, more work is needed in the future aiming to develop novel contributions to the area, by developing guidelines and methods for embedding penalties as well from the start in the design of a data-sharing framework.

## VI. TESTING AND EVALUATION

We evaluated the performance of the proposed framework by measuring the transaction cost and time used by all the methods to change the contract's states as shown in Fig. 4 and Table 1. The monetary cost of the system mainly comes from the gas cost of operating smart contracts on the Ethereum platform. Fig. 4 shows one instance of the transaction cost for the execution of the smart contract function.

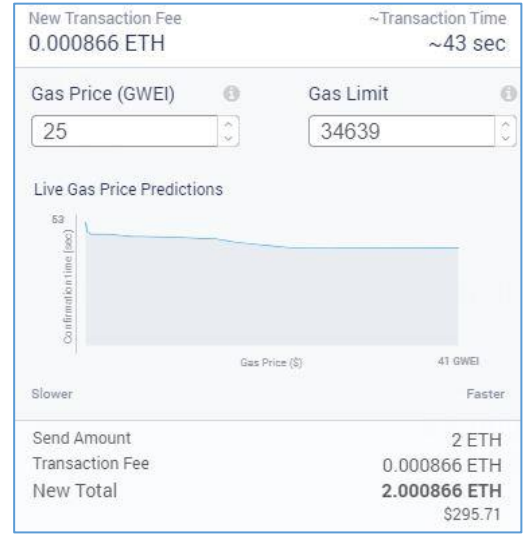

Fig. 4. Tx cost for executing a smart contract function.

TABLE I. TX COST FOR EXECUTING DIFFERENT SMART CONTRACT FUNCTIONS.

| Functions | ContractState | Cost (Gas) | Tx fee (ETH) | Tx time (sec) |
|---|---|---|---|---|
| contractDeploy() | Created | 834625 | 0.020866 | < 46 |
| consumerPay() | Locked | 34639 | 0.000866 | ~ 43 |
| paymentSettle() | Inactive | 47611 | 0.001247 | ~ 43 |

The evaluation involved deploying the codes with the Remix IDE [37] on the Ethereum Ropsten testnet [38]. Gas Station [39] provides three categories of gas price: SafeLow (less than 30 minutes), Standard (less than 5 minutes) and Fast (less than 2 minutes). The gas limit is helpful to optimize the gas usage in order to provide a safety mechanism, as sometimes code with bugs might keep on consuming unnecessary gas for the execution.

We carried out the experiment to obtain the gas costs and currency equivalents for all contract function calls, based on the standard practice with a fixed gas price of 25 Gwei (2.5*1e-8 ETH) and a gas limit 34639 Gwei. The price of ETH on March 25, 2019, was $177.88. The nodes responded quickly in all our cases with a befitting transaction cost as shown in Table 1.

We recorded the costs of all the contract function calls including the cost for executing the constructor. The cost associated with the contract deployment (calling constructor function) for the Gas Price of 25 Gwei was 0.020866 ETH and the total cost of all function calls was 0.002113 ETH ($0.375). The cost to execute function1 to change the contract state from "Created" to "Locked" was lesser than that to execute function2 to change the state from "Locked" to "Inactive". This difference is justified because the funtion2 caused the contract to transfer ether to both the data consumer and provider, whereas the function1 only cased the data consumer to transfer ether to the contract. Therefore, the data consumers will gradually be overwhelmed with the increase in the total cost since they pay more for accessing the data.

## VII. CONCLUSION

We presented a new framework which is based on user-controlled privacy and data-sharing policies encoded in smart contracts. It supports building up a verifiable record of the provenance, accountability of access and incentives for users to share their research data, in terms of rewards (micro-payments or credits) through blockchain. It reduces the risk of leaking of the user data through data anonymization with AES encryption technique. It also complies with the GDPR by not storing any of the user profile into the blockchain. It provides the provision of off-chain storage of any published data with their hashes and meta-data committed into the blockchain and the rapid delivery of the data over the P2P network. In this way, users become owners of their data and can decide how their data is collected and used, as well as shared. We also presented a review of related works on research data sharing, our proposed blockchain-based framework and an evaluation of the framework by measuring the transaction cost for smart contracts deployment. The results show that nodes responded quickly in all tested cases with a befitting transaction cost. In future work, we will conduct user surveys to study the usefulness and usability of the approach, and the trust users can have in the system because user studies are much needed to investigate the users' perception on the acceptance of our proposed solution and the effects of different variables of the behavioral models on the actual use of the system.